# Universal and ultrafast quantum computation based on free-electron–polariton blockade

Aviv Karnieli[1], Shai Tsesses[2,3], Renwen Yu[4], Nicholas Rivera[5], Ady Arie[6], Ido Kaminer[2] and Shanhui Fan[4]

[1]Raymond and Beverly Sackler School of Physics and Astronomy, Tel Aviv University, Ramat Aviv 69978, Tel Aviv, Israel
[2]Andrew and Erna Viterbi department of Electrical and Computer Engineering, Technion – Israel Institute of Technology, Haifa 32000, Israel
[3]Department of Physics and Research Laboratory of Electronics, Massachusetts Institute of Technology, Cambridge, MA 02139, USA
[4]Department of Electrical Engineering, Stanford University, Stanford, California 94305, USA
[5]Department of Physics, Harvard University, Cambridge, MA 02138, USA
[6]School of Electrical Engineering, Fleischman Faculty of Engineering, Tel Aviv University 69978, Tel Aviv, Israel

**Abstract**

Cavity quantum electrodynamics (QED), wherein a quantum emitter is coupled to electromagnetic cavity modes, is a powerful platform for implementing quantum sensors, memories, and networks. However, due to the *fundamental* tradeoff between gate fidelity and execution time, as well as limited scalability, the use of cavity-QED for quantum computation was overtaken by other architectures. Here, we introduce a new element into cavity-QED – a free charged particle, acting as a flying qubit. Using free electrons as a specific example, we demonstrate that our approach enables ultrafast, deterministic and universal discrete-variable quantum computation in a cavity-QED-based architecture, with potentially improved scalability. Our proposal hinges on a novel excitation blockade mechanism in a resonant interaction between a free-electron and a cavity polariton. This nonlinear interaction is faster by several orders of magnitude with respect to current photon-based cavity-QED gates, enjoys wide tunability and can demonstrate fidelities close to unity. Furthermore, our scheme is ubiquitous to any cavity nonlinearity, either due to light-matter coupling as in the Jaynes-Cummings model or due to photon-photon interactions as in a Kerr-type many-body system. In addition to promising advancements in cavity-QED quantum computation, our approach paves the way towards ultrafast and deterministic generation of highly-entangled photonic graph states and is applicable to other quantum technologies involving cavity-QED.

**Introduction**

Quantum computation has the potential to unlock unprecedented capabilities for solving difficult problems[1], and has been the focal point of research across many physical disciplines, with several realizations thus far[2–10]. One of the earliest approaches to quantum computation relies on cavity quantum electrodynamics (cavity-QED)[11–17]: a system in which a stationary matter qubit (such as an atom[18–22], a quantum dot[23–29], a defect center[30] or even a superconducting artificial atom[31–33]) interacts strongly with a "flying" (i.e., propagating) photonic qubit[18,34–36] via the engineering of their electromagnetic environment (e.g., via a cavity)[18,20,23,34,37–43]. Quantum computing in CQED[13,18–21,26,28,29,33,35,36,44–47] can be achieved by applying quantum gates through cavity-mediated interactions between the matter qubits[44,47], the photonic qubits[18,21], or a combination thereof[35,36,48,49].

Though CQED remains an important platform for several quantum technologies, including quantum sensing[50,51] and quantum networks[34,44,48,49,52–55], other architectures for quantum computing[2–10] are currently far more mature. One of the fundamental issues plaguing cavity-QED as a quantum computing platform is its inherent tradeoff between gate fidelity and speed[45,56,57]. The better the cavity quality and emitter coherence, the more time it requires for interaction between the matter and photonic qubits[45,56,57], making fast, large bandwidth and high-fidelity gates impractical, especially when attempting to implement cavity-QED with inherently scalable platforms, such as integrated photonic circuits[24,58–60]. Scalability of cavity-QED especially in a solid-state environment is also hindered by the need to fabricate a large number of indistinguishable emitters in identical cavities, which is difficult given the inherent variance in cavity resonance and emitter transitions that can hamper gate fidelities.

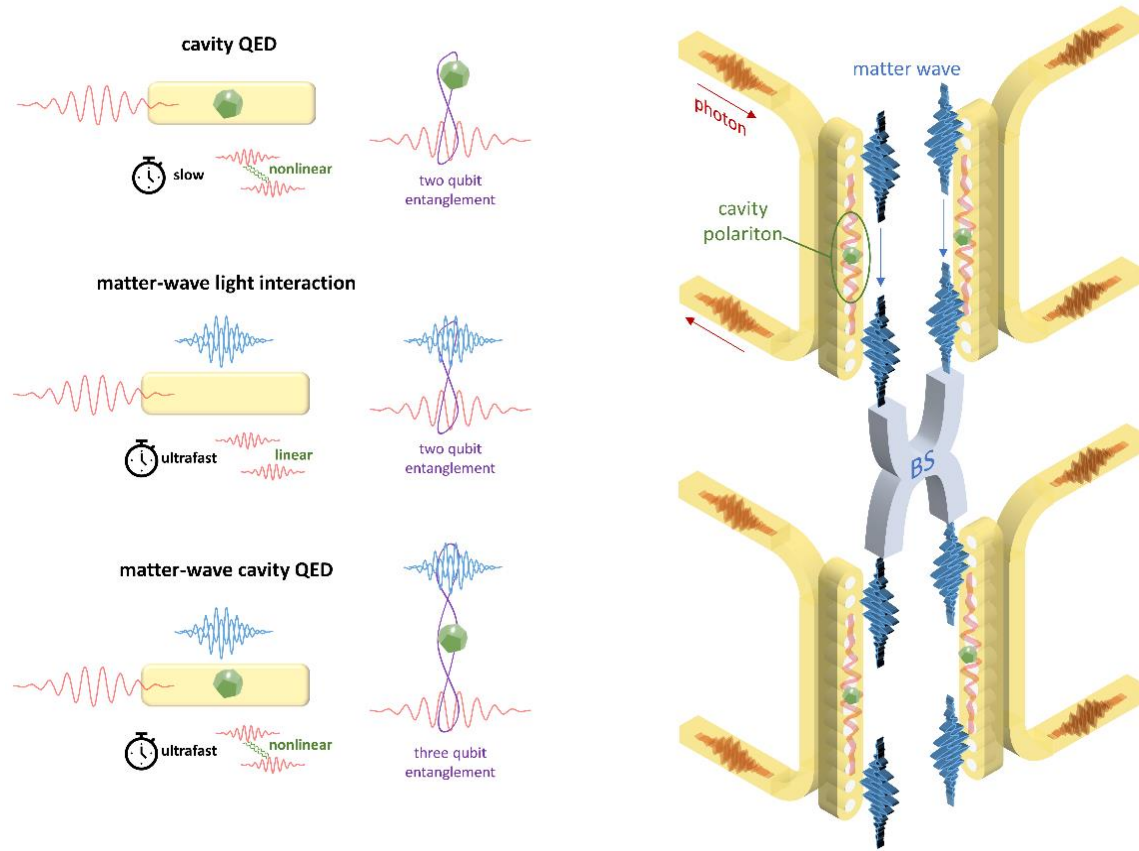

Fig. 1: **Integrating flying charged particle qubits into cavity-QED**. **a** In standard cavity QED, a stationary matter qubit interacts nonlinearly with a flying photonic qubit, in an interaction mediated by a cavity. Light-matter entanglement between the flying and stationary qubits is generated over slow time scales. **b** A flying charged particle qubit (such as an electron or ion) interacts linearly with a photonic qubit on an ultrafast timescale, resulting in entanglement between them. **c** When a flying charged particle qubit drives a cavity-QED system, the light-matter interaction can be both nonlinear and ultrafast, resulting in three-qubit entanglement between the constituents of the system. **d** Concept illustration of cavity-QED implemented in integrated photonic circuits and driven by a flying charged particle qubit: ultrafast quantum gates and entanglement distribution between separate cavity polaritons is mediated by the path- and energy-encoded information of the flying charged particle qubit (manipulated by matter wave beam splitters). Each excited cavity polariton comprises an entangled light-matter state, and its photonic part can still be used as a flying qubit, for long-distance communication between processors or over a quantum network.

In this work, we propose to incorporate a new type of qubit into cavity-QED – a flying charged particle qubit, based on propagating charged quantum particles such as free electrons and ions, as shown in Fig. 1. Focusing on free electrons, we identify a mechanism for a *free-electron–polariton blockade*, which allows for the deterministic generation of a single hybrid light-matter excitation (a polariton), and its entanglement with a free electron, solely through the near-field interaction of the electron and the cavity. Using this free-electron–polariton blockade, we construct a universal set of quantum gates between polaritons for deterministic, discrete-variable quantum computation. For cavity-QED systems possessing strong enough nonlinearity, we show that these quantum gates can operate at timescales of less than a picosecond – orders of magnitude faster than the current state-of-the-art in cavity-QED, and significantly below the dissipation rate of the cavity-QED system. This ultrafast time scale is made possible owing to a resonant phase matching between the free-electron and the cavity nearfield, and need only be longer than the nonlinearity timescale (e.g., the vacuum Rabi oscillation in the cavity). Our proposal is compatible with other cavity-QED model systems, and greatly relaxes the tradeoff between gate speed and fidelity. The flying electron qubit is inherently tunable through its kinetic energy, robust to variations in cavity resonance and immune to absorption, unlike a flying photon qubit. Thus, it can boost quantum information capacity and provide a new avenue to scale up cavity-QED-based architectures. Our scheme even facilitates the deterministic generation of highly-entangled multi-photon states[61–63], and can potentially upgrade all manner of quantum technologies based on CQED, ranging from quantum sensing[64] to quantum simulation[65].

### **Theoretical model: free-electron–polariton blockade**

The quantum interaction between free electrons and light has seen dramatic advancements in the past decade, from the observation of photon-induced near-field electron microscopy[66–69] to the emerging field of free-electron quantum optics[70–81], which is fully compatible with integrated photonics[82,83]. These include demonstrations of electron-photon correlations[83–85], highly-efficient quantum coupling between free electrons and photons[86], and theoretical proposals for continuous-variable quantum optical computation[87]. Complementarily, the quantum interaction between free electrons and quantum emitters gave rise to several theoretical investigations[88–93] focusing on the ability to infer and manipulate the emitter state.

However, we are still at the early stages of exploring the interaction of free-electrons with *nonlinear* systems, which include coupling between light and matter[64,94,95] or between different light fields[96,97]. Thus, a fully quantum theoretical description of such systems is necessary, which we provide below, beginning with an intuitive explanation.

When a quantum free-electron resonantly interacts with a linear cavity, it can efficiently couple to the optical mode and spontaneously emit multiple photons[70,74,86], as illustrated in Fig. 2a. In the course of the interaction, the electron energy becomes entangled with the generated photon number. In contrast, when a free electron is resonantly coupled to a cavity-QED-like system, it interacts with a *nonlinear* cavity possessing an inherent polariton blockade mechanism: the existence of a first polariton in the cavity detunes the excitation of a second polariton. Hence, the free electron can only emit a single polariton, as illustrated in Fig. 2b. The electron becomes entangled only with this single excitation, and the system therefore occupies a much smaller Hilbert space befitting discrete-variable computation.

Our model incorporates free-electrons with a cavity-QED system, and captures all the effects mentioned above. The total system Hamiltonian is given by

$$H_{\text{tot}} = \underbrace{H_{\text{p}} + H_{\text{matt}} + H_{\text{e}}}_{H_0} + \underbrace{H_{\text{nl}} + H_{\text{ep}}}_{H_1}, \quad (1)$$

where $H_0 = H_{\text{p}} + H_{\text{matt}} + H_{\text{e}}$ is the non-interacting part of the Hamiltonian; $H_{\text{e}} = E + \int dq\, \hbar v q c_{k+q}^\dagger c_{k+q}$ is the linearized free-electron Hamiltonian under the paraxial approximation[70], with $E, v, k$ being the initial energy, velocity and wavenumber of the electron, respectively, and $c_k$ being the fermionic annihilation operator; $H_{\text{p}} = \hbar \omega a^\dagger a$ is the photonic Hamiltonian, describing a mode with energy $\hbar\omega$ and annihilation operator $a$; and $H_{\text{matt}}$ is the stationary emitter Hamiltonian, which is model-dependent.

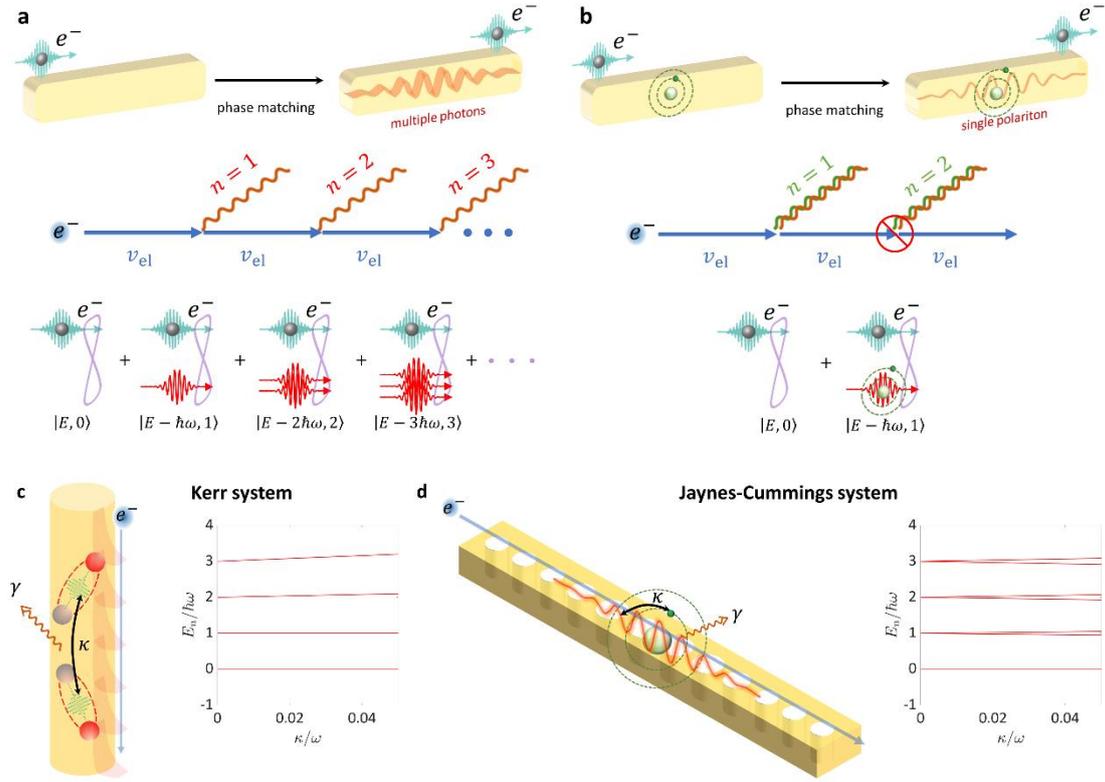

Fig. 2: **Free-electron–polariton blockade**. **a** A quantum free electron traverses a linear cavity while being phase-matched to the cavity mode, conserving both momentum and energy. If the interaction is strong enough, the electron is able to emit multiple photons into the cavity, and its energy becomes highly entangled with the cavity photon number. **b** When the same electron traverses a cavity-QED-like system, which is comprised of a nonlinear cavity, the inherent polariton blockade effect allows for phase matching only in the process of a single polariton emission, while consecutive emission events will be phase-mismatched. As a result, the electron energy becomes entangled in a minimal polariton-number Hilbert space. The dynamics then reduces to an effective two-level system driven by a quantum free-electron. **c,d** cavity-QED models for a free-electron polariton blockade. In both settings, the electron grazes the cavity in vacuum and interacts with the polariton nearfield. The first model is a Kerr nonlinear system (**c**), characterized by the interaction Hamiltonian $H_{\text{nl}} = \hbar\kappa a^\dagger a^\dagger aa$ and realized by interacting exciton polaritons with creation and annihilation operators $a^\dagger, a$ in microcavities or waveguides[98–101]. Inset: the polaritonic energy levels as a function of the nonlinearity ratio $\kappa/\omega$, with spacing linearly increasing with polariton number $n$. The second model is a Jaynes-Cummings system (**d**), characterized by the interaction Hamiltonian $H_{\text{nl}} = \hbar\kappa\sigma_+ a + \hbar\kappa\sigma_- a^\dagger$ and realized, for example, by situating a two-level system with ladder operators $\sigma_+, \sigma_-$ inside a nanobeam single-mode microcavity[23,24,102] with photonic creation and annihilation operators $a^\dagger, a$. Inset: polaritonic energy as a function of $\kappa/\omega$. While the Rabi splitting increases as $\sqrt{n}$, the spacing between adjacent energy levels decreases as $\sqrt{n} - \sqrt{n-1}$.

The interaction part of the Hamiltonian, $H_1 = H_{\text{nl}} + H_{\text{ep}}$, is comprised of the emitter-photon and electron-photon interactions, respectively. The former Hamiltonian depends on the specific cavity-QED model of choice, while the latter is given in the interaction picture, and under the nonrecoil approximation[103], by the expression

$$H_{\text{ep}} = i\hbar \int dq\, e^{i(qv-\omega)t} g_q b_q^\dagger a - i\hbar \int dq\, e^{-i(qv-\omega)t} g_q^* b_q a^\dagger, \quad (2)$$

where $q$ is the longitudinal recoil experienced by the free-electron; $b_q = \int dk\, c_{k-q}^\dagger c_k$ is the free-electron momentum-lowering operator by a recoil $q$, satisfying $\left[b_q, b_{q'}^\dagger\right] = 0$; and $g_q = (ev/\hbar\omega)\int_0^L e^{-iqz}\mathcal{E}_z(\mathbf{r}_T, z)dz$ is the recoil-dependent coupling constant, with $e, \mathcal{E}(\mathbf{r}_T, z), \mathbf{r}_T, L$ being the electron charge, the cavity mode envelope, the transverse position of the electron inside the cavity nearfield and the cavity length, respectively.

The nonlinear cavity Hamiltonian $H_{\mathrm{nl}}$ is generally quantified by a nonlinearity strength $\kappa$, stemming from strong coupling between light and matter. The essential difference between a nonlinear and a linear cavity is the emergence of a number-dependent energy difference between adjacent energy levels. Instead of a constant difference, i.e. $\omega_n - \omega_{n-1} = \omega$, a nonlinear cavity exhibits the relation $\omega_n - \omega_{n-1} = \omega + f(n)\kappa$, where $f(n)$ is some function of the energy level number $n$. This number dependence facilitates the polariton blockade mechanism. In Appendix A, we specify two typical models for cavity-QED-like nonlinear cavity systems, determining both $H_{\mathrm{matt}}$ and $H_{\mathrm{nl}}$: the Jaynes-Cummings (JC) model[16,17,23,24,34], considering a single two-level emitter inside a cavity; and a Kerr nonlinearity model, with inherent single-photon nonlinearity that is mediated by matter-matter interactions in a many-body system[98–101]. These two systems and their nonlinear energy spectra are depicted in Fig. 2c,d.

In the presence of photonic losses (radiative or absorptive, with a total rate $\gamma$), which we assume to be the dominant loss mechanism in our system, the interaction-picture dynamics (subscript $I$) of the system density matrix, $\rho_I$, is given by the Lindblad master equation of the form

$$\frac{d\rho_I}{dt} = -\frac{i}{\hbar}[H_{1,I}, \rho_I] + \gamma\left(a\rho_I a^\dagger - \frac{1}{2}a^\dagger a\rho_I - \frac{1}{2}\rho_I a^\dagger a\right), \quad (3)$$

where $H_{1,I}$ is the interaction-picture form of $H_1 = H_{\mathrm{nl}} + H_{\mathrm{ep}}$ where $H_{\mathrm{nl}}$ depends on our model of choice (see Appendix A and Fig. 2c-d). Unless stated otherwise, we will numerically integrate Eq. (3) to obtain all the results presented in this work.

In the limit of resonant phase-matching ($L \gg 2\pi/q_0$, where $q_0$ is the longitudinal wavenumber of the cavity mode) and negligible losses ($\gamma \ll v/L$), it is possible to invoke energy and momentum conservation in the excitation of the $n$-th energy level, such that $q = q_0$, $b_q = b_{q_0} \equiv b$, $g_q = g\delta(q - q_0)$ and $q_0 v = \omega_n - \omega_{n-1}$. The scattering matrix associated with the interaction thus takes an analytical closed form, which depends on the properties of the cavity. In the linear case, it takes the well-known form of a displacement operator[70,74], i.e., $S_{\mathrm{lin}} = \exp(g_Q b^\dagger a - g_Q^* b a^\dagger)$, where $g_Q = gL/v$ is the dimensionless coupling constant. In contrast, the scattering matrix for the nonlinear cavity takes a distinctly different form, given by

$$S_{\mathrm{nl}} = \cos|\Omega| - i\sin|\Omega|\left(e^{i\arg\Omega}b|\bar{1}\rangle\langle\bar{0}| + e^{-i\arg\Omega}b^\dagger|\bar{0}\rangle\langle\bar{1}|\right), \quad (4)$$

where the states $|\bar{n}\rangle$, $n = 0,1$ denote any pair of two consecutive polaritonic number states on the polariton energy ladder (detailed in the caption of Fig. 3), and $\Omega$ is the free-electron–polariton vacuum Rabi frequency, both of which are model-dependent ($\Omega = g_Q$ or $g_Q/\sqrt{2}$ for the Kerr and JC models, respectively; see Appendix B). Contrary to the linear case, where spontaneous emission of the electron into the cavity results in Poissonian statistics with a mean $|g_Q|^2$, a strongly nonlinear cavity supports only a single polariton excitation, and can produce a complete population inversion between the ground state (or, in general, the lower energy state of the pair) $|\bar{0}\rangle$, and the consecutive excited state $|\bar{1}\rangle$, which we refer to as the free-electron–polariton blockade effect.

Figs. 3a,b illustrate the continuous transition between the linear and nonlinear regimes with increasing $\kappa$, as is evidenced in both the photon statistics and the electron energy loss spectrum (EELS) (see Appendix C for the calculation method). Figs. 3c,d, on the other hand, demonstrate the sharp dependence of the blockade effect on the phase-matching condition, whereby a blockade occurs only at the exact phase-matching condition with a specific polariton energy level. Nevertheless, the phase-matching bandwidth – being inversely proportional to the electron time of flight across the cavity, $v/L$ – can tolerate small variations in cavity resonance (say, on the order of the cavity linewidth), which can otherwise detune a photonic flying qubit.

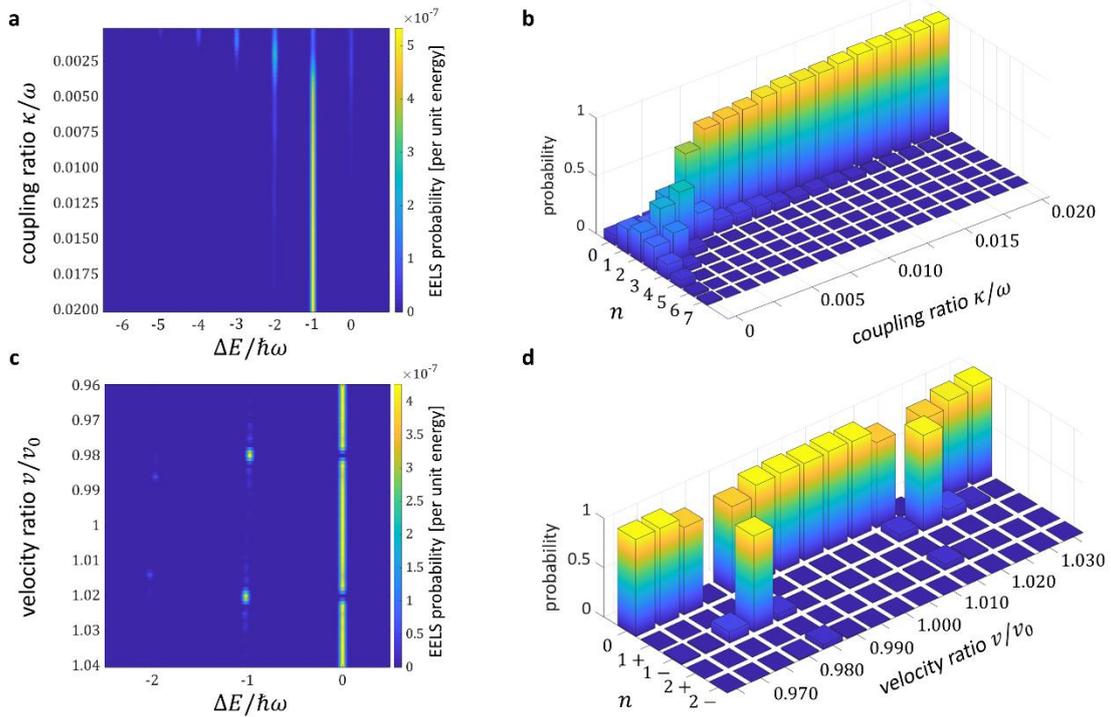

Fig. 3: **Polariton statistics and electron energy loss spectra of the free-electron–polariton blockade: dependence on cavity nonlinearity and phase-matching selectivity**. Simulation results of electron energy loss spectra (EELS) and polariton number statistics, via the numerical integration of the Lindblad master equation [Eq. (3)]. **a,b** Continuous transformation of the EELS (**a**) and photon number statistics (**b**) for free-electron spontaneous emission into a Kerr nonlinear cavity as a function of the coupling ratio $\kappa/\omega$. As the nonlinearity increases, complete population inversion from the Kerr-polariton ground state $|\bar{0}\rangle = |0\rangle$ and a single Kerr-polariton Fock state $|\bar{1}\rangle = |1\rangle$ occurs, whereas a Poissonian distribution is apparent in the linear case. Parameters used in the simulation are $E = 200$ keV, $L = 40\mu m$, $\gamma/\omega = 10^{-5}$, $v = \omega/q_0 = 0.6953c$, $q_0 = 2\pi/532$ nm$^{-1}$, $g_Q = \pi/2$. **c,d** EELS (**c**) and polariton number statistics (**d**) for free-electron spontaneous emission into a Jaynes-Cummings nonlinear cavity as a function of the velocity ratio $v/v_0$ (where $v_0 = \omega/q_0 = 0.2719c$). $\kappa/\omega = 0.02$, $g_Q = \pi/\sqrt{2}$, $E = 20$ keV and all other parameters are the same as in (**a,b**). In the JC example, we consider the single excitation manifold of $|\bar{0}\rangle = |0^*\rangle = |0, g\rangle$ and $|\bar{1}\rangle$, which can be either the upper ($|1+\rangle$) or lower ($|1-\rangle$) polariton state, where $|1\pm\rangle = (|g, 1\rangle \pm |e, 0\rangle)/\sqrt{2}$, with $|0\rangle, |1\rangle$ denoting the bare photon number states and $|g\rangle$ ($|e\rangle$) the bare emitter ground (excited) state. Only when the free-electron velocity is tuned to phase match with a specific polariton [$v = v_\pm = (\omega \pm \kappa)/q_0 = (1 \pm 0.02)v_0$], does a full population inversion take place. In this manner, the electron velocity can be used to selectively determine the polariton type chosen for excitation. The difference in $g_Q$ between (**a,b**) and (**c,d**) arises from our assumption that in the JC model, the electron only couples to the photonic degrees of freedom of the cavity.

In a specific phase matching condition, the blockade reduces the entire Hilbert space to that of an effective two-level system with Rabi oscillation determined by $|g_Q|$, as shown in Fig. 4. Interestingly, tuning the electron velocity allows for selective excitation of different polaritonic transitions, which could become useful for encoding different physical qubits and for generating higher-order Fock states.

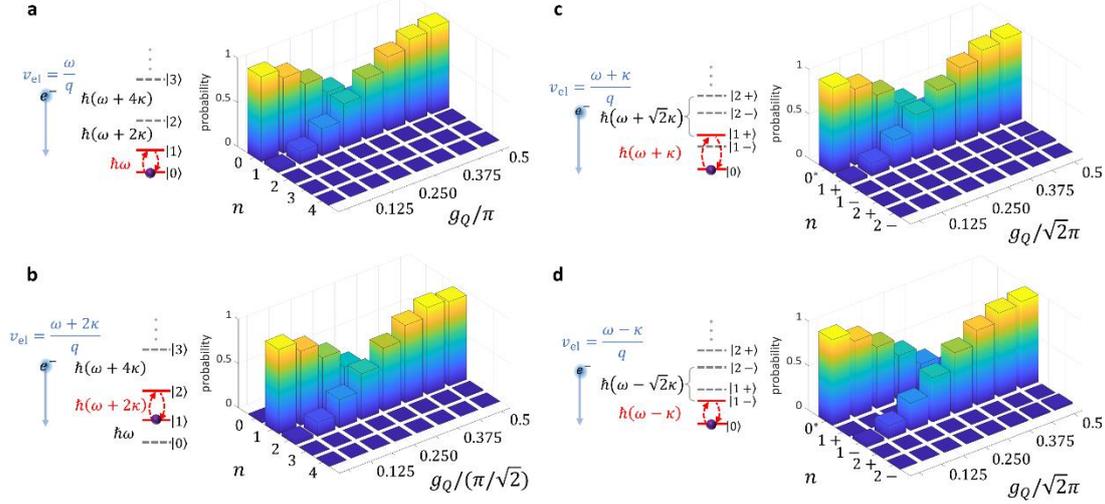

Fig. 4: **Effective photonic two-level system dynamics driven by free electrons.** Polariton number statistics for both the Kerr (**a,b**) and JC (**c,d**) models, as a function of the dimensionless electron-light coupling $g_Q$. A single Rabi oscillation is apparent in all figures within a manifold pre-selected by the electron velocity, producing phase-matching with a specific polariton transition [**a**: $v = \omega/q$, **b**: $v = (\omega + 2\kappa)/q$, **c**: $v = (\omega + \kappa)/q$, **d**: $v = (\omega - \kappa)/q$]. Each transition is illustrated in the corresponding inset nonlinear cavity energy level diagram. Stimulated emission in the cavity produces a faster population inversion in (**b**), which occurs at $g_Q = \pi/(2\sqrt{2})$. The coupling ratio $\kappa/\omega$ was set to 0.02 in all simulations. Otherwise, all parameters in (**a,b**) and (**c,d**) are the same as in Fig.3(**a,b**) and (**c,d**), respectively.

To verify that the polariton blockade can indeed be used to control the quantum state of polaritons and produce quantum gates, we calculate the fidelity $\mathcal{F} = \langle\psi|\rho|\psi\rangle$ between the state of the full system [described by the density matrix $\rho$ in Eq. (3)], and the target pure state $|\psi\rangle$ generated by the ideal scattering matrix [Eq. (4)], for different combinations of relative loss $\gamma/\omega$ and coupling $\kappa/\omega$. Figs. 5a,b show the fidelity map assuming either the JC (Fig. 5a) or Kerr (Fig. 5b) model, demonstrating a clear universal behavior. Fig. 5c presents a comparison of fidelities at the strongest coupling coefficient considered ($\kappa/\omega = 0.02$), as a function of loss ratio, where a maximal fidelity is shown to exceed 97% for both models. We emphasize that this is not a fundamental limit, as considering slower electrons[104,105], longer electron-cavity interaction length (which has been achieved experimentally[86,106]), or even multiple passes using a free-electron cavity[107], can increase the fidelity indefinitely, and can also be used to reach high fidelity values for weaker nonlinearities.

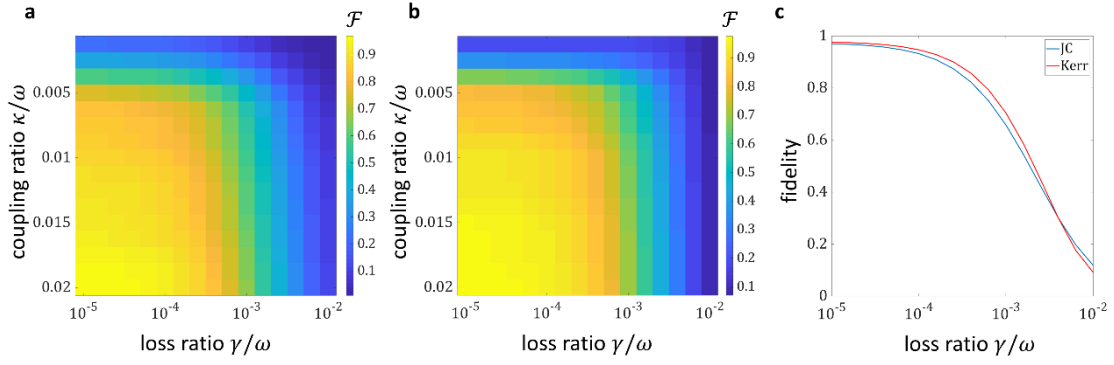

Fig. 5: **Quantum state fidelity as a function of coupling strength and cavity loss**. Quantum state fidelity for different photon loss rates and coupling strengths, calculated between the total density matrix $\rho_I$ of Eq. (3) and the target state generated by the analytic scattering matrix of Eq. (4). **a** Jaynes-Cummings cavity, with parameters $E = 20$ keV, $v = (\omega - \kappa)/q_0 = 0.2719c$, $q_0 = 2\pi/532$ nm$^{-1}$, $L = 40$μm and $g_Q = \pi/\sqrt{2}$. **b** Kerr cavity, with parameters $E = 200$ keV, $v = \omega/q_0 = 0.6953c$, $q_0 = 2\pi/532$ nm$^{-1}$, $L = 40$μm and $g_Q = \pi/2$. **c** Comparison of the polariton state fidelity in each model, for the maximal coupling $\kappa/\omega = 0.02$. Maximal fidelity for both models exceeds 97%. Apart from the cavity loss rate, the limiting factor for the fidelities is the interplay between the cavity nonlinearity $\kappa$ and the phase-matching bandwidth, proportional to $v/L$. If the latter is not much smaller than the former, the electron couples to higher-order polariton states that lie outside of the target Hilbert space.

**Universal, ultrafast free-electron-mediated quantum gates**

Even without a free electron present, both of the models we considered can produce quantum gates between their constituents to perform discrete-variable quantum computation, using interactions controlled and mediated by the nonlinear cavity. However, they suffer from significant limitations to gate fidelity and speed, owing to the need for large coupling efficiencies between the facilitator of the interaction (i.e., cavity) and the interacting parties (i.e., photons and/or emitter). This is not merely a technical issue, since the requirements for spectral and spatial matching between modes outside and inside the cavity impose an upper boundary to the rate at which quantum gates may be applied[45,56,57], typically on the order of 1 MHz (or once every 1 microsecond).

Therein lies the advantage of integrating electrons into these models: as an electron passes by a cavity, its interaction with the cavity persists as long as it interacts with the cavity near-field. Not only does this eliminate the need for mode matching, since the electron inserts a photon into the cavity with greater ease, but it also restricts the gate application rate to the interaction time between the electron and the cavity. Considering cavities on the order of tens of microns, and fast electrons, such as can be found in transmission electron microscopes, gates can be applied within hundreds of femtoseconds (or a rate of a few THz – 6 orders of magnitude in difference).

The first step towards integrating free-electrons with cavity-QED is to identify the physical qubits that best fit the proposed interaction mechanism. Unlike conventional cavity-QED protocols employing the JC model, wherein the emitter qubit is typically encoded in two long-lived ground states[108], here it is more natural to encode qubits in the single-excitation manifold of the polaritons (the polariton-number basis $|\bar{0}\rangle$ and $|\bar{1}\rangle$). While this encoding would be impractical for conventional cavity-QED gates in quantum memory and network applications that require long lifetimes, in our case the computation can be done orders of magnitude faster than the polariton lifetime, rendering this encoding useful for the quantum gate operations we discuss below. After the computation is performed, the tunable free-electron–polariton blockade mechanism (Fig. 4) can be employed to convert the qubit encoding from the polariton-number basis to other bases (for example, as in Fig. 6c), allowing for a seamless integration with conventional cavity-QED protocols.

Next, for the notion of using free-electron-mediated quantum gates in a cavity-QED-like architecture, we must demonstrate that they can support a universal set of quantum gates. Building on the analytical description of the free-electron–polariton blockade in Eq. (4), we proceed to denote the electron-cavity interaction by a single unitary operator $U(\Omega) = S_{\mathrm{nl}}$, where $S_{\mathrm{nl}}$ is the nonlinear scattering matrix of Eq. (4). We will use the unitary $U(\Omega)$ to construct a universal single-qubit gate set acting on the polariton qubit driven by a free electron ancilla. However, as the electron energy can become entangled with the polariton state, care must be taken to make sure that the former *does not* change due to the applied gate. In other words, the free-electron energy must be *disentangled* from the polariton qubit.

First, we can describe an electron-controlled rotation by an angle $\phi$ about the z-axis of the polariton state Bloch sphere, taking the form $C_{\mathrm{ep}}R_z(\phi) = |E\rangle_0\langle E|_0 \otimes I_{\mathrm{p}} + |E\rangle_1\langle E|_1 \otimes \exp(-i\phi\sigma_z)$, as illustrated in Fig. 6a. Here, the electron is a path qubit denoted as $|E\rangle_m$, where $m = 0,1$ represents an electron passing far away from the cavity ($m = 0$) or in close proximity to it ($m = 1$). If the electron interacts twice with the same cavity but at different transverse locations ($a$ and $b$) within the complex cavity nearfield envelope, having the same interaction strength $|\Omega_a| = |\Omega_b| = \pi/2$ and a relative phase $\phi = \arg\Omega_b - \arg\Omega_a$, the electron energy before and after interaction remains the same. The resulting polariton gate becomes

$$R_z(\phi) = \exp(-i\phi\sigma_z) = U\left(\frac{\pi}{2}e^{i\phi}\right)U\left(\frac{\pi}{2}\right) = e^{-i\phi}|\bar{0}\rangle\langle\bar{0}| + e^{i\phi}|\bar{1}\rangle\langle\bar{1}|, \quad (5)$$

where we used the fact that $bb^\dagger = b^\dagger b = 1$. Discarding a global phase, the gate $R_z(\phi)$ is the general phase gate $P(2\phi)$, and for $\phi = \pi/2, \pi/4$ and $\pi/8$, we can implement electron-controlled $Z, S$ and $T$ gates, respectively.

Additionally, we describe an electron-controlled rotation by an angle $|\Omega|$ about a transverse axis $\hat{\boldsymbol{\rho}} = (\cos\varphi, \sin\varphi)$ of the polariton state Bloch sphere, taking the form $C_{\mathrm{ep}}R_{\hat{\boldsymbol{\rho}}}(|\Omega|) = |\mathrm{comb}(\varphi)\rangle_0\langle\mathrm{comb}(\varphi)|_0 \otimes I_{\mathrm{p}} + |\mathrm{comb}(\varphi)\rangle_1\langle\mathrm{comb}(\varphi)|_1 \otimes \exp(-i|\Omega|\hat{\boldsymbol{\rho}} \cdot \boldsymbol{\sigma}_T)$, as illustrated in Fig. 6b. Here, the electron is prepared in a free-electron comb state[74] of phase $\varphi$, ideally defined as $|\mathrm{comb}(\varphi)\rangle = \sum_l e^{il\varphi}|E + l\hbar\omega_0\rangle$. Being eigenstates of the electron ladder operators $b$ and $b^\dagger$, electron comb states do not get entangled with the polariton qubit. Thus, applying the unitary in Eq. (4) on the state $|\mathrm{comb}(\varphi)\rangle_1$ (and assuming that $\arg\Omega = 0$ for simplicity), we find

$$\begin{aligned}R_{\hat{\boldsymbol{\rho}}}(|\Omega|) = \exp(-i|\Omega|\hat{\boldsymbol{\rho}} \cdot \boldsymbol{\sigma}_T) &= U(|\Omega|)|\mathrm{comb}(\varphi)\rangle_1 \\ &= \left[\cos|\Omega| - i\sin|\Omega|\left(e^{i\varphi}|\bar{1}\rangle\langle\bar{0}| + e^{-i\varphi}|\bar{0}\rangle\langle\bar{1}|\right)\right] \otimes |\mathrm{comb}(\varphi)\rangle_1,\end{aligned} \quad (6)$$

In this manner, we can implement elementary gates such as $\exp(-i(\pi/2)\sigma_x) = -iX$, $\exp(-i(\pi/4)\sigma_y) = HZ$ and the Hadamard gate $\exp(-i(\pi/2)\sigma_x)\exp(-i(\pi/4)\sigma_y) = -iH$. Finally, combining the two electron-controlled gates of Eqs. (5) and (6) implements a universal set of single-qubit gates, which can be readily used for state preparation and readout done exclusively by free electrons. The gate $C_{\mathrm{ep}}R_{\hat{x}}$, as presented in Fig. 6c, can serve a purpose other than performing single-qubit operations: by selectively tuning the free-electron–polariton blockade to a different polaritonic transition, it can transform a physical qubit in the polariton number basis to other bases, such as the polariton-type (upper and lower polariton) basis, or to a qubit encoding using two emitter ground states. This transformation may be important for certain cavity-QED applications such as quantum memories and quantum networks[34,44,48,49,52–55].

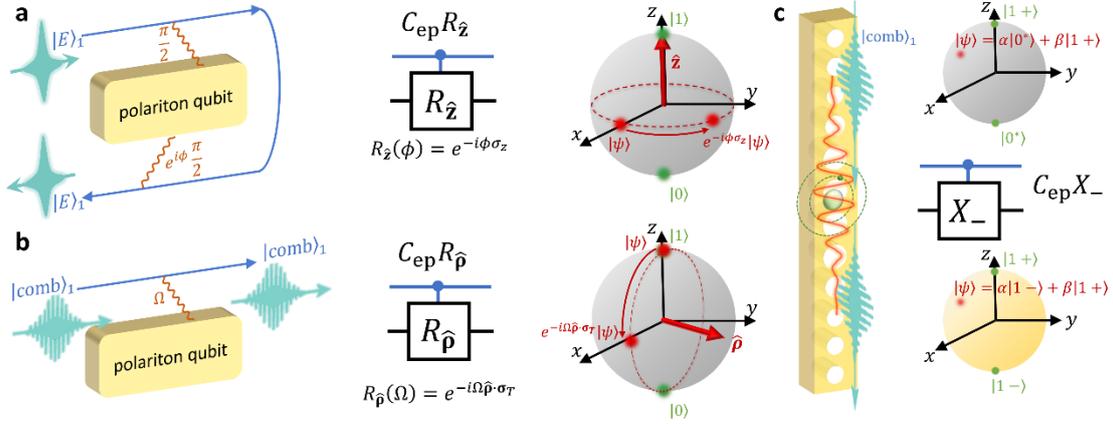

Fig. 6: **Universal free-electron-driven single-qubit gates for the polariton state**. a Electron-controlled z-rotation on the polariton Bloch sphere, implemented by letting an electron interact twice with the same cavity but at different transverse locations within the nearfield. The coupling has the same magnitude $|\Omega| = \pi/2$ for each pass, yet the two paths can have a phase difference $\phi$. b Electron-controlled transverse rotation gate about a general transverse axis $\hat{\rho} = (\cos\varphi, \sin\varphi)$ on the polariton bloch sphere, implemented by preparing an electron in a comb state $|\text{comb}(\varphi)\rangle$ and letting it interact with the cavity. c Transformation of the physical qubit basis from the number basis $\{|0^*\rangle, |1+\rangle\}$ (top inset) to the polariton-type basis $\{|1-\rangle, |1+\rangle\}$ (bottom inset), by applying the gate $C_{ep}X_-$ with a comb electron, having a velocity tuned with the transition $|0^*\rangle \leftrightarrow |1-\rangle$.

Now that we have established a universal single-qubit gate set, we must also present a two-qubit gate between polaritons that supports universal quantum computation, with the electron acting as an ancilla[109]. A deterministic two-polariton gate can be achieved by the interaction of an electron with two consecutive cavities, while making sure to disentangle the electron ancilla from the state of the two polariton qubits.

The first stage of a proposed two-polariton controlled-$Z$ gate is an entanglement operation wherein the first polariton qubit controls the path of the free electron, realizing a *controlled path* ($C_{pe}\text{Path}$) gate, as depicted in Fig. 6a. First, the free electron ancilla qubit is prepared in state $|E\rangle_1$ and interacts with the first cavity with coupling $\Omega = (\pi/2)e^{i\phi}$, entangling the electron energies $|E \pm \hbar\omega\rangle_1$ with the polariton computational basis states. Next, the electron passes through an electron spectrometer, separating the two electron energies into two different paths. Finally, a second interaction with the cavity, with the same coupling $\Omega = (\pi/2)e^{i\phi}$, restores the initial electron energy $E$ in each of the output paths. Acting on a general polariton qubit state $|\psi\rangle = \alpha|\bar{0}\rangle + \beta|\bar{1}\rangle$ and a free electron ancilla in state $|E\rangle_1$, the $C_{pe}\text{Path}$ gate results in

$$C_{pe}\text{Path} |\psi\rangle|E\rangle_1 = \alpha|0\rangle|E\rangle_0 + \beta|1\rangle|E\rangle_1 = X_e C_{pe}X|\psi\rangle|E\rangle_1, \qquad (7)$$

where $X_e$ and $C_{pe}X$ denote, respectively, a NOT gate acting on the electron and a CNOT gate with polariton control qubit and electron target qubit. The equivalent circuit of $C_{pe}\text{Path}$ is depicted in Fig. 7a.

The next stage of the gate would be to apply a controlled-$Z$ gate on the second polariton qubit, controlled by the free electron (the $C_{ep}Z$ gate of Fig. 6a). Next, a Hadamard gate $H$ is applied on the electron qubit. The latter can be implemented for example using ponderomotive beam splitters[110], elastic Kapitza-Dirac diffraction from a standing wave[111], or a PINEM interaction[67], as depicted in Fig. 7b. Finally, a second $C_{ep}Z$ gate is applied, this time on the first polariton qubit. The Hadamard and second $C_{ep}Z$ gate disentangle the free-electron from the polariton qubits, after which the electron is found deterministically in the state $(|E\rangle_0 + |E\rangle_1)/\sqrt{2}$. A final application of a Hadamard gate transforms the ancilla state to $|E\rangle_0$. Fig. 7c depicts the explicit and equivalent circuit of the two-polariton CZ gate (see Fig. 8 in Appendix D for derivation). Clearly, the action of the equivalent circuit is to implement a controlled-$Z$ operation between the two bottom qubits, mediated by the top ancilla qubit.

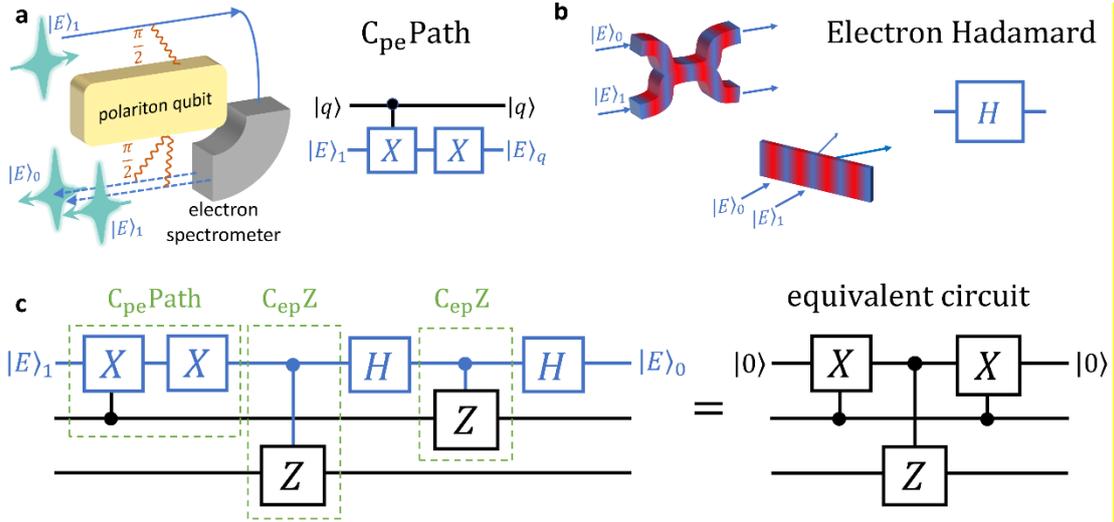

Fig. 7: **Two-polariton controlled-$Z$ gate mediated by a free-electron**. a controlled-path ($C_{pe}$Path) gate between a polariton control qubit and a free electron target qubit, prepared in an initial state $|E\rangle_1$. The electron interacts with the cavity with coupling strength $\Omega = \pi/2$, either losing or gaining an amount $\hbar\omega$ of energy. Then, the electron passes through an electron spectrometer which translates the resulting different electron energies into two different output paths, both interacting again with the same cavity, restoring the original electron energy. The result is an entangled state wherein the polariton controls the output path of the electron, with an equivalent circuit as shown in the inset. b A free-electron Hadamard gate (realized, for example, by a ponderomotive beam splitter, Kapitza-Dirac elastic diffraction, a PINEM interaction, etc.). c The quantum circuit of the two-polariton controlled-$Z$ gate and its equivalent form. The first polariton qubit controls the free electron ancilla by $C_{pe}$Path. Then, the electron drives a $C_{ep}Z$ gate acting on the second polariton qubit. The electron is disentangled from the polaritons by application of an electron Hadamard gate $H$ followed by a second $C_{ep}Z$ acting on the first qubit, and a final Hadamrd, leaving the ancilla qubit in state $|E\rangle_0$. In the equivalent circuit (derived in Fig. 8 in Appendix D), it is clear that a controlled-$Z$ gate is applied to the two bottom qubits, mediated by the top ancilla qubit.

## Discussion

We proposed a new paradigm for quantum computation based on a cavity-QED-like architecture with flying charged particle qubits mediating interactions, providing a potentially enormous increase in gate application speed. We specifically focused on free-electron flying qubits interacting with cavities hosting single-photon nonlinearities of different types (JC and Kerr), introducing a novel free-electron–polariton blockade effect. The blockade reduces the entire Hilbert space to that of an effective two-level system driven by the free electron (similarly to the on-resonance AC Stark effect[112]), establishing that it can indeed enable discrete-variable gate-based quantum computation. We then described a universal quantum gate set in cavity-QED-like architectures, enabled solely by free electrons.

Contrary to the prevalent photon-based gates, which need to operate on timescales of the order of the cavity lifetime to maintain high fidelity, free-electron–based gates can operate as fast as the time-of-flight $T = L/v$ of the electron, only limited by the Rabi oscillation time within the cavity. Therefore, these gates will always be fundamentally faster and more robust to errors in the cavity parameters, such as detuning of the order of the cavity linewidth [while such changes might affect the cavity nonlinearity itself in some CQED models (JC), others will be indifferent to them (Kerr)].

Furthermore, the electron can interact with several cavities without being absorbed, unlike the photon[49,53], and its inherent coherent energy width $\Delta E$ along with the phase-matching bandwidth of its interaction with the cavity mode can aid in tolerating small changes to the resonance of different cavities while retaining high gate fidelities. Therefore, introducing even a single flying charged particle qubit can aid in the speed and scalability of cavity-QED quantum computing, whereas adding more flying charged particle qubits can increase both the dimension of the computation[113–115] and the Hilbert space dimension of the system.

Moreover, we note that our proposal can be potentially integrated in different cavity-QED quantum computing architectures, as using the flying charged particle qubit does not impede existing methods for producing quantum gates, only adding upon them. In fact, photon-based quantum gates are more preferable for long range communications between quantum processors or within a quantum network[34,44,48,49,52–55], and thus the current strengths of the cavity-QED platform can still be maintained. Moreover, unlike earlier demonstrations of cavity-QED with flying neutral atom qubits[116–118], the free-electron–light interaction relies on a fundamentally different mechanism[70], establishing the free charged particle qubit as a genuinely new entity in the cavity-QED model.

The demonstration of our proposed architecture for quantum computation should motivate further improvement in the performance of the state-of-the-art in cavity-QED systems and in achieving strong free-electron-photon coupling. The most important requirement for realizing the full potential of our proposal is to find a cavity-QED system possessing a strong enough nonlinearity, which is challenging in photonic microcavity systems. The requirement for a highly-efficient electron-photon interaction ($g_Q$ larger than unity) is a further challenge. More specifically, the high blockade fidelity sets an upper bound for the application rate of the electron-mediated quantum gates, since the blockade frequency – on the order of the cavity nonlinearity, $\kappa$ – must be far greater than the free-electron phase-matching bandwidth $\Delta\omega_{PM} = 1/T = v/L$. In turn, this relative bandwidth $\Delta\omega_{PM}/\omega$ must be far greater than the relative energy uncertainties of both the electron ($\Delta E/E$) and the cavity mode ($\gamma/\omega$). Thus, in general, we require

$$\frac{\gamma}{\omega}, \frac{\Delta E}{E} \ll \frac{\Delta\omega_{PM}}{\omega} \ll \frac{\kappa}{\omega}. \quad (8)$$

The range of parameters currently demonstrated in experiment suggests that a proof-of-principle experiment of the free-electron polariton blockade effect can be pursued. State of the art values of $\kappa/\omega$ are $10^{-4}$ with quantum dots in dielectric microcavities[23,24,119], $10^{-3} \mu m^2$ (per unit density) in Kerr-like exciton polariton systems[99–101], and $10^{-8}$ in atomic CQED systems[34,53]. In the optical domain, an efficient electron-photon coupling $g_Q$ of unity with $\Delta\omega_{PM}/\omega$ of $7 \times 10^{-4}$ has already been achieved[86] (and could readily be made smaller), and the typical values of $\gamma/\omega, \Delta E/E$ (between $10^{-4}$ to $10^{-9}$, depending on quality factors and electron energies), are indeed much smaller than that, as required.

However, to fully utilize the advantage of the flying charged particle qubit for quantum computation, significant improvements in the single-photon nonlinearity and the electron-photon coupling in all-dielectric cavities are still necessary. Quantum dots in nanobeam microcavities[23,24] with record-low mode volumes[120], and propagating dipolar exciton polaritons in waveguides[99–101] are promising platforms to increase the cavity nonlinearity. Using slower electrons[104,105,121] or electron cavities[107] could support lower nonlinearity values, albeit at the cost of slower gates. On the other hand, increasing the electron-photon coupling is possible using integrated photonic cavities, based on slot waveguides[122], nanowires[123,124], flatband photonic modes[125], photonic crystals on nanobeams[120] and fiber Bragg gratings supporting BICs[126]. Interestingly, the stringent requirements of Eq. (8) could potentially be fulfilled in the microwave range of circuit QED systems[127], by employing few-keV electrons and centimeter-length transmission line resonators, operating in the ultrastrong coupling regime[128,129].

The interaction of a single electron with several cavities can be used to deterministically generate large-scale photonic graph states[61–63] with high fidelities, which is one of the premier bottlenecks for photonic measurement-based quantum computation[130,131]. In the same vein, continuous-variable, fault-tolerant quantum computation can also be performed using similar systems to those considered here[87,132,133], implying that flying charged particle qubits might also enable quantum error correction in cavity-QED-based platforms[32,132,134] as well.

Finally, we stress that the free-electron–polariton blockade effect and its consequences, as detailed in our work, can open a myriad of other applications in all manner of quantum technologies. Aside from enhancing cavity-QED-based quantum sensing[50], free electrons can act as mediators for interactions between different cavities in quantum simulations of the Jaynes-Cummings-Hubbard model[65]. Potentially, this would allow much larger hopping amplitudes, opening a route towards using CQED systems to simulate Heisenberg interaction models[135], as well as interaction Hamiltonians with controllable dimensionality, sign and symmetry through time-varying electron acceleration and deceleration between consecutive interactions.

**Acknowledgements**

A.K. and S.T. acknowledge support by the Adams Fellowship of the Israeli Academy of Sciences and Humanities. S.T. further acknowledges support from the Yad Hanadiv Foundation through the Rothschild fellowship and from the Israeli Council for Higher Education. S. F and R. Y. acknowledge the support of the ACHIP Collaboration funded by the Gordon and Betty Moore Foundation (GBMF4744). A. A. acknowledges the support of the Israel Science Foundation, grant no 969/22. N.R. acknowledges the support of a Junior Fellowship from the Harvard Society of Fellows.

**Appendix A**

In this Appendix, we detail the two CQED models considered in this work. The first is a Kerr-type cavity, having an intrinsic nonlinearity stemming from an effective interaction between the hybrid light-matter polaritonic eigenstates of the system. The system usually consists of an ensemble of emitters inside the optical cavity, such that the Hopfield model[16] can be invoked, effectively describing the polaritons as bosons with annihilation operator $a$. Usually, only the polaritons in the lower branch are considered[98], such that the noninteracting part of the Hamiltonian is simply given by a linear harmonic oscillator

$$H_\text{p} + H_\text{matt} = \hbar \omega a^\dagger a, \quad (A1)$$

An effective interaction between the polaritons (mediated by their matter parts) dictates a Kerr-like Hamiltonian, given as $H_\text{nl} = \hbar \kappa a^\dagger a^\dagger a a$, (A2) which could be realized, for example, by considering microcavities hosting interacting exciton-polaritons[98–101], as depicted in Fig. 2c. The system has polaritonic eigenstates $|n\rangle$ having frequencies $\omega_n = n\omega + n(n-1)\kappa$, and a frequency spacing that increases with excitation number, i.e., $\omega_n - \omega_{n-1} = \omega + 2(n-1)\kappa$.

The second model we consider is a Jaynes-Cummings (JC) type nonlinearity[16,17,23,24,34], which results from the strong coupling between two-level emitters and optical cavity modes[16], depicted in Fig. 2d. The noninteracting part of the system Hamiltonian is

$$H_\text{p} + H_\text{matt} = \hbar \omega a^\dagger a + \frac{\hbar \omega}{2}\sigma_z, \quad (A3)$$

while the interaction is dominated by the Hamiltonian

$$H_\text{nl} = \hbar \kappa \sigma_+ a + \hbar \kappa \sigma_- a^\dagger, \quad (A4)$$

where $\sigma_-, \sigma_+, \sigma_z$ are the ladder and Pauli $z$ operators, respectively, of the single two-level emitter embedded in the microcavity (such as an atom, molecule, or quantum dot). The emitter is assumed to be tuned to the cavity resonance, and can be found either in its ground state $|g\rangle$ or an excited state $|e\rangle$. The has both upper and lower polaritonic eigenstates, $|n+\rangle = (|g,n\rangle + |e,n-1\rangle)/\sqrt{2}$ and $|n-\rangle = (|g,n\rangle - |e,n-1\rangle)/\sqrt{2}$, in addition to a ground state $|0^*\rangle = |g0\rangle$. The polariton excitation frequencies are, respectively, $\omega_{n\pm} = n\omega \pm \sqrt{n}\kappa$, and the frequency spacing between polaritons of the same type decreases with excitation number, i.e., $\omega_{n,\pm} - \omega_{n-1,\pm} = \omega \pm (\sqrt{n} - \sqrt{n-1})\kappa$. Note that in the electron interaction Hamiltonian of Eq. (2), we neglected the coupling between the electron and the emitter, since it is usually much weaker[88–90].

## Appendix B

In this Appendix, we derive the expression for the analytic form of the scattering matrix. We first exemplify the free-electron polariton blockade effect for the case of a Kerr cavity. Momentum conservation introduces a sharply-peaked electron-photon coupling, taking the limiting form $g_q \to g\delta(q - q_0)$, which, upon performing the $dq$ integral in Eq. (1), expanding the operators $a$ and $a^\dagger$ in terms of the nonlinear cavity eigenstates (which for a Kerr cavity, are simply Fock states), and tuning the electron velocity such that $q_0 v - \omega = 0$ (i.e., only electrons of $q = q_0 = \omega/v$ contribute), reduces the electron-photon Hamiltonian to

$$H_\text{ep} = \sum_{n=1}^{\infty} e^{i2(n-1)\kappa t} i\hbar g b^\dagger \sqrt{n}|n-1\rangle\langle n| + h.c., \quad (B1)$$

where $b \equiv b_{q_0}$, and where [unlike Eq. (2) and (3)] we have considered an interaction picture wherein $H_{\text{nl}}$ is part of the noninteracting Hamiltonian $H_0$, and $H_1 = H_{\text{ep}}$. For the JC model, we write the photon number states in terms of the JC eigenstates as $|n, g\rangle = (|n +\rangle + |n -\rangle)/\sqrt{2}$, $|0, g\rangle = |0^*\rangle$ and $|n, e\rangle = (|n + 1, +\rangle - |n + 1, -\rangle)/\sqrt{2}$, and again expand the photon ladder operator $a$ in terms of these eigenstates. Tuning the electron velocity to the first upper-polariton excitation (the transition $|0^*\rangle \to |1 +\rangle$) such that $q_0 v - \omega - \kappa = 0$, we find

$$H_{\text{ep}} = \frac{1}{\sqrt{2}} i\hbar g b^\dagger \left( |0^*\rangle\langle 1 +| + e^{-i2\kappa t} \frac{1}{\sqrt{2}} |0^*\rangle\langle 1 -| \right)$$
$$+ \frac{1}{2} i\hbar g b^\dagger \sum_{n=1}^{\infty} f_{n+} e^{i(f_{n-}-1)\kappa t} |n, +\rangle\langle n + 1, +|$$
$$+ f_{n+} e^{-i(f_{n-}+1)\kappa t} |n, -\rangle\langle n + 1, -| + f_{n-} e^{-i(f_{n+}+1)\kappa t} |n, +\rangle\langle n + 1, -|$$
$$+ f_{n-} e^{i(f_{n+}-1)\kappa t} |n, -\rangle\langle n + 1, +| + \text{h.c.}, \quad (B2)$$

where $f_{n\pm} = \sqrt{n+1} \pm \sqrt{n}$. In the linear limit of both models, $\kappa = 0$ and all photonic excitations are perfectly tuned with the electron. When nonlinearity is present, the excitations other than the resonant transition of choice are detuned by an amount of at least $2\kappa$ (Kerr model) or $(2 - \sqrt{2})\kappa$ (JC). These transitions accumulate a dynamical phase that rapidly oscillates when $\kappa T \gg \pi$, where $T = L/v$ is the interaction time of the electron with the cavity. In this limit of strong nonlinearity, we can adopt the rotating-wave approximation by dropping these rapidly oscillating terms in the Hamiltonian, to finally obtain

$$H_{\text{ep}} = i\hbar \begin{cases} g b^\dagger a - g^* b a^\dagger, & \text{linear} \\ g b^\dagger |0\rangle\langle 1| - g^* b |1\rangle\langle 0|, & \text{Kerr} \\ \frac{g}{\sqrt{2}} b^\dagger |0^*\rangle\langle 1 +| - \frac{g^*}{\sqrt{2}} b |1 +\rangle\langle 0^*|, & \text{JC} \end{cases}, \quad (B3)$$

We note that similar Hamiltonians can be obtained for any two consecutive transitions on the polariton energy ladder, with an adequate tuning of the electron velocity, as shown in Fig. 4.

The scattering matrix is in general given by $S = \mathcal{T} \exp\left[-\frac{i}{\hbar} \int_{-\infty}^{\infty} dt H_{\text{ep}}(t)\right]$, where $\mathcal{T}$ stands for time-ordering. As $H_{\text{ep}}$ is now time-independent, $S$ can assume a closed form given by

$$S = \begin{cases} \exp(g_Q b^\dagger a - g_Q^* b a^\dagger), & \text{linear} \\ \cos|g_Q| - i \sin|g_Q| \left(e^{i \arg g_Q} b|1\rangle\langle 0| + e^{-i \arg g_Q} b^\dagger |0\rangle\langle 1|\right), & \text{Kerr} \\ \cos\frac{|g_Q|}{\sqrt{2}} - i \sin\frac{|g_Q|}{\sqrt{2}} \left(e^{i \arg g_Q} b|1 +\rangle\langle 0^*| + e^{-i \arg g_Q} b^\dagger |0^*\rangle\langle 1 +|\right), & \text{JC} \end{cases}, \quad (B4)$$

where $g_Q = gL/v$ is the dimensionless coupling constant. To capture the universal behavior of both CQED models, we define the polariton physical qubits $|\bar{0}\rangle$ and $|\bar{1}\rangle$ of the single-excitation manifold, and denote by $\Omega = g_Q$ or $g_Q/\sqrt{2}$ the electron-photon coupling in the Kerr and JC models, respectively, recovering the expression for the nonlinear scattering matrix of Eq. (4).

**Appendix C**

In this appendix, we present the derivations of the electron energy loss spectrum (EELS) and polariton number statistics derived from the numerically-simulated density matrix of Eq. (3). The density matrix of the full system is given by

$$\rho = \begin{cases} \int dk \int dk' \sum_{n} \sum_{n'} \rho(n,k;n',k') |n,k\rangle\langle n',k'|, & \text{Kerr} \\ \int dk \int dk' \sum_{n} \sum_{n'} \sum_{s} \sum_{s'} \rho(n,s,k;n',s',k') |n,s,k\rangle\langle n',s',k'|, & \text{JC} \end{cases}, \quad (C1)$$

where $k, n, s$ denote the electron wave-vector, photon number, and atomic state ($s = e, g$), respectively. To obtain the EELS spectrum, we trace out the photonic (and in the JC model, also the atomic) degrees of freedom, resulting in the reduced electron density matrix $\rho_{el} = \text{Tr}_{ph,a}\{\rho\}$. The EELS spectrum is given by the diagonal elements of $\rho_{el}$, expressed in terms of the total density matrix $\rho$ as

$$p_{EELS}(k) = \langle k|\rho_{el}|k\rangle = \begin{cases} \sum_{n} \rho(n,k;n,k), & \text{Kerr} \\ \sum_{n} \sum_{s} \rho(n,s,k;n,s,k), & \text{JC} \end{cases}, \quad (C2)$$

To obtain the polariton number statistics, we apply a similar procedure, this time tracing-out the electron degrees of freedom, resulting in the photon (or joint photon-atom) density matrix $\rho_{ph,a} = \text{Tr}_{el}\rho$. The polariton number statistics is given by the diagonal elements of $\rho_{ph}$ in the Kerr model, and by the diagonal elements of $\rho_{pol} = U\rho_{ph,a}U^\dagger$, where $U$ is a unitary transformation performing a change of basis to the JC polariton eigenstates. The result in terms of the total density matrix $\rho$ is

$$P(n) = \langle n|\rho_{ph}|n\rangle = \int dk \rho(n,k;n,k), \quad (C3)$$

for the Kerr model and

$$P(n\pm) = \langle n\pm|\rho_{pol}|n\pm\rangle = \sum_{ms} \sum_{m's'} U_{n\pm,ms} U^*_{m's',n\pm} \int dk \rho(m,s,k;m',s',k), \quad (C4)$$

for the JC model.

**Appendix D**

This appendix contains a graphical derivation of the quantum circuit equivalent to the two-polariton controlled-Z interaction of Fig. 6c.

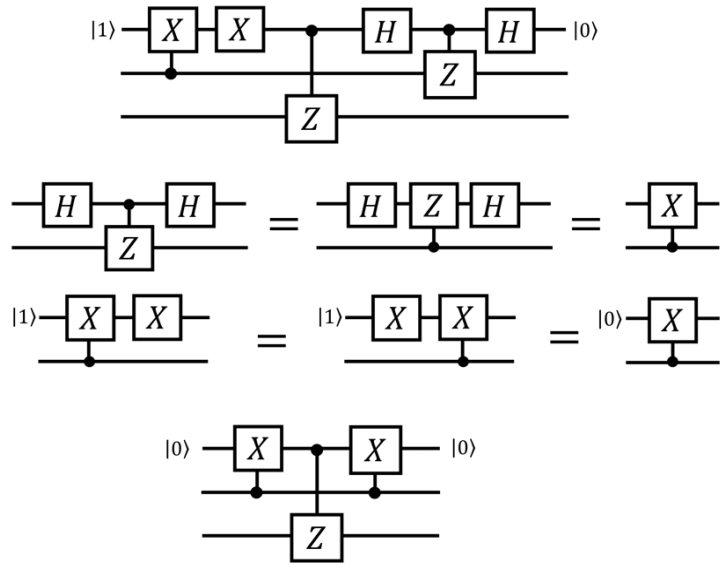

Fig. 8: equivalent quantum circuit to the two-polariton controlled-Z gate mediated by an electron ancilla qubit.